\documentclass[twocolumn,showpacs,showkeys,preprintnumbers,amsmath,amssymb]{revtex4}

\usepackage{graphicx}
\usepackage{dcolumn}
\usepackage{bm}

\begin{document}

\title{Kirchhoff's Loop Law and   the maximum   entropy production principle } 

\author{Pa\v{s}ko \v{Z}upanovi\'{c}}
 \email{pasko@pmfst.hr}
\author{Davor Jureti\'{c}}
 \email{juretic@pmfst.hr}
\affiliation
{Faculty of Natural Sciences, Mathematics and Education \\ University of Split, Teslina 12,  21000 Split \\ Croatia}
\author{Sre\'{c}ko Botri\'{c}}
\email{srecko.botric@fesb.hr}
\affiliation{ Faculty of Electrical Engineering, Mechanical Engineering and Naval Architecture \\ University of Split, Bo\v{s}kovi\'{c}eva b.b.  21000 Split \\ Croatia} 

\date{\today}

\begin{abstract}
  In contrast to the standard derivation of  Kirchhoff's loop law, which invokes 
 electric  potential, we  show, for the linear planar electric network
in a  stationary state at the fixed temperature,
 that  loop law can be derived from   
  the maximum   entropy production principle.  This means that
  the currents  in network branches  are distributed   in such a way    as
to achieve the state of   maximum  entropy  production.
\end{abstract}

\maketitle

\section{\label{1} Introduction}

 Kirchhoff's laws \cite{Kirchhoff} are the standard part 
   of  general physics courses 
 \cite{feyman,zemansky,harold}. In electrical engineering they are the starting point for  the 
  analysis of   stationary  processes in electric networks \cite{lePage}. In the   stationary state, due to 
  the principle of  charge conservation,   Kirchhoff's current law (current law) is valid. It states that in 
 each node of the network the sum of ingoing currents equals the sum of outgoing currents.
  Kirchhoff' loop law (loop law) is based on the assumption that 
 electric potential is a well defined  quantity  in any point of the electric network.
 Then one can apply the principle of  
 energy conservation to a macroscopic small amount of the charge circulating around the loop, i.e. the energy obtained
 on the sources should be equal to the dissipated energy. This statement is equivalent to the loop law, which states that 
the algebraic sum of electromotive forces (EMF-s) of the sources is equal to the sum of voltages (potential differences) in
 the   loop.
In this paper we show that  the loop law  can be derived for a linear planar network using  
  the maximum  entropy production principle \cite{dewar}. This means that   
  stationary state currents  distribute themselves in the branches in such a way  as to maximize the  entropy production in the network.

\section{\label{2} Mesh currents and loop law}

We consider a planar network (see Fig. 1).
 \begin{figure}
\resizebox{8cm}{8cm}
 {\includegraphics{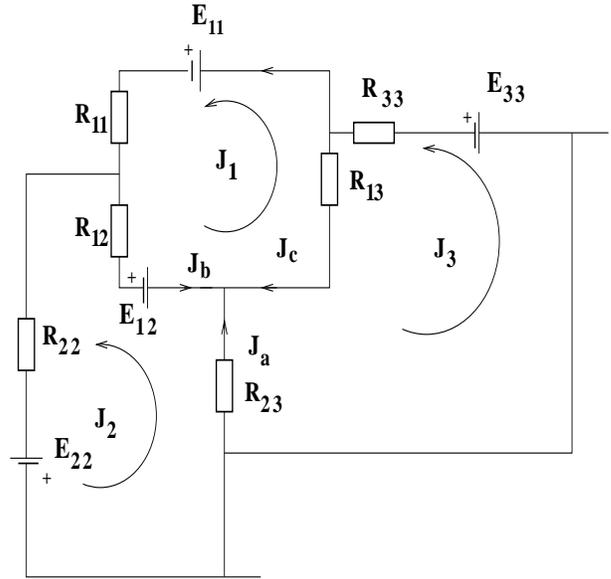}}
\caption{\label{fig1} A linear planar electric network}
\end{figure}
Assuming that network parameters, EMF-s and resistances, are fixed, one can find all currents  applying current and 
 loop law to  nodes and loops. However, due to the current law, the currents in branches are not independent quantities. 

 \begin{figure}[h]
\resizebox{7cm}{7cm}
{\includegraphics{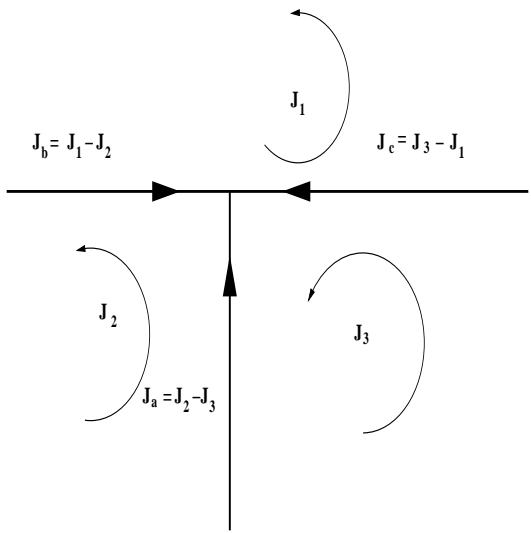}}
\caption{\label{fig2} Mesh currents and  Kirchhoff's current law}
\end{figure}

Kirchhoff's laws  give no prescription on how to find a set of
 independent currents for a given electric network.
In the case of the  planar network this problem has been solved by electrical engineers \cite{lePage}  by
 introducing the concept of mesh currents.
Let us first define, within the  network, the  simple loop  (mesh) as the  one  having no loop within it
 (loops 1,2 and 3 in Fig.\ref{fig1}).
We associate   a mesh current with each mesh 
  ($J_1,J_2$ and $J_3$ in Fig. \ref{fig1}). A current in a branch,  common to  two neighboring meshes, is  an algebraic sum
 of    corresponding mesh currents (see Fig.\ref{fig2}). The current in the outer branch, the  branch which belongs to one mesh  only,
 is equal to the  mesh current. Evidently the mesh currents incorporate the current law (see Fig.\ref{fig2}).  It is easy to  prove,
 by means of the mathematical induction, that   the number of contour
 currents  is equal to 
 the number of independent currents in the network. 
The mesh currents are independent parameters determining the stationary state of the electric  network as the thermodynamic system.
In order to make the analysis of the network in terms of the mesh currents as simple as possible we introduce  equivalent EMF-s
  and 
 resistances.
 The equivalent EMF
 is  equal to the algebraic sum of the
 EMF-s in a certain branch and the equivalent  resistance is the  sum of the  resistances in that branch.
 We enumerate the  meshes and corresponding mesh currents by single index notation, while the double index notations is
 used for the equivalent EMF-s and for equivalent resistances (Fig.\ref{fig1}). 
Different indices in the double index notation appear when the single branch is shared between two meshes.

Applying the loop law for each mesh loop in terms of  mesh currents we obtain the system of linear equations,  
 the number of 
 equations being equal to the number of mesh currents,
\begin{equation}
\label{lin}
\sum_j E_{ij}=R_{ii}J_i + \sum_j R_{ij}(J_i-J_j).
\end{equation}
Note that the system of equations (\ref{lin}) incorporates both Kirchhoff's laws (due to the definition of the mesh currents).
Here the left hand side of the equation is the algebraic sum of EMF-s within $i-th$ mesh loop. The EMF is positive if  mesh 
 current comes out of its positive pole. Otherwise it is negative. Evidently it holds   true
\begin{equation}
\label{anti}
E_{ij}=-E_{ji} \;\;\;  i \neq j,
\end{equation}
and
\begin{equation}
\label{comu}
R_{ij}=R_{ji}.
\end{equation}
 We stress  that the  system of equations (\ref{lin})
 assumes that the  principle of  energy conservation holds for each mesh separately (loop law). In the following we show 
that this system of  equations (\ref{lin}) can be alternatively  derived applying  the maximum  entropy 
 production  principle under the condition
 that the  principle of  energy conservation is valid for the whole  network as the thermodynamic system.

\section{\label{3} Conservation of the energy}

 If the system at the fixed absolute  temperature $T$ releases  heat per  unit of time, $dQ/dt$, the corresponding  
 entropy production is
\begin{equation}
\label{enpr}
\frac{d_iS}{dt}= \frac{1}{T} \frac{dQ}{dt},
\end{equation}
where index $i$, as it is  introduced in the theory of  non-equilibrium processes \cite{prigogine},
 emphasizes that only  irreversible entropy change (rise) is considered. 

   The entropy production is the function of the  thermodynamic state of a system.
The stationary state of   resistor at the fixed temperature  is
 defined by the magnitude    and the sense of   the flow of the 
 electric current. Due to the isotropy of the resistor the entropy production 
 is invariant with respect to the flow direction of the current,
\begin{equation}
\frac{dS_i}{dt}(-J)=\frac{dS_i}{dt}(J).
\label{J-J}
\end{equation}
 Assuming that the stationary state of the resistor is close to its equilibrium state 
 Taylor expansion of the entropy production gives,
\begin{equation}
\frac{dS_i}{dt}=a+bJ+cJ^2+dJ^3 + {\cal O}(J^4).
\label{exp}
\end{equation}
A free  term  vanish  since it corresponds to the entropy production in the equilibrium state.
Due to the relation (\ref{J-J}) coefficients of all odd powers vanish 
 while 
$ c > 0$  due to
  $ dS_i/dt >0$.
Putting   $c=R/T$, where $R$ is the resistance 
  we get 
 the well known expression for the dissipated electric power $dQ/dt=RJ^2$.
Evidently  $c>0$ implies $R>0$.
 
    A linear planar network can be separated into sources of the forces (EMF-s) and the passive part (resistors).  
 In the stationary state  there is no change in internal energy of the resistors while they  
 convert energy from the sources into the heat  given off to  the surroundings.
 The rate of the 
 energy conversion is given by,
\begin{equation}
\label{Q}
\frac{dQ}{dt}= \sum_i R_{ii}J_i^2+ \frac{1}{2} \sum_{ij}  R_{ij}(J_i-J_j)^2.
\end{equation}
Energy released by all sources (EMF-s) per unit of time is 
\begin{equation}
\label{W}
\frac{dW}{dt}=
\sum_i E_{ii}J_i +\frac{1}{2}\sum_{ij} E_{ij}(J_i-J_j)
\end{equation}
The factor $\frac{1}{2}$, in these equations,  appears due to the double counting in the case of  internal branches.

 Bearing in mind that resistors do not change their internal energy we can write, according to  the first law of thermodynamics,
\begin{equation}
\label{1law}
\frac{dQ}{dt}-\frac{dW}{dt}=0,
\end{equation}
i.e. taking into account (\ref{Q}) and (\ref{W})
it holds, 
\begin{eqnarray}
\label{enconv}
\sum_i E_{ii}J_i +\frac{1}{2}\sum_{ij} E_{ij}(J_i-J_j)= \nonumber \\
= \sum_i R_{ii}J_i^2+ \frac{1}{2} \sum_{ij}  R_{ij}(J_i-J_j)^2.
\end{eqnarray}

\section{\label{4} Entropy production and its extremum in the electric network}

We argue that Kirchhoff's loop law follows from the  principle of  the maximum   overall  entropy production in the network, assuming that
 the energy conservation (\ref{enconv}) is satisfied.
If there are  $n$ meshes, and $n$  associated 
 mesh  currents, we have a conditional  extremum problem in the  $n$-dimensional linear  space.

Assuming that  all resistors in the network are at the same temperature $T$ the maximum of entropy production 
 occurs at the same point of the $n-$dimensional  space of currents  
 as the   maximum of generated heat (see eq. \ref{enpr}).

Standard procedure \cite{krasnov} of solving the conditional extremum introduces  Lagrange's multipliers. 
In this case one seeks the extremum of the function 
\begin{equation}
\label{stand}
F=\frac{dQ}{dt}+ \lambda \Psi,
\end{equation}
 where due to the condition (\ref{enconv})
\begin{eqnarray}
\label{enconv1}
\Psi =\sum_j E_{jj}J_j +\frac{1}{2}\sum_{i,j} E_{ij}(J_i-J_j)- \nonumber \\ 
-\left[\sum_i R_{ii}J_i^2+ \frac{1}{2} \sum_{i,j}  R_{ij}(J_i-J_j)^2\right]=0.
\end{eqnarray}
The function $F$ now reads
\begin{eqnarray}
\label{Psi}
F=(1-\lambda)\left[ \sum_i R_{ii}J_i^2+ \frac{1}{2} \sum_{i,j}  R_{ij}(J_i-J_j)^2 \right]+ \nonumber \\
+\lambda \left[ \sum_i E_{ii}J_i +\frac{1}{2}\sum_{i,j} E_{ij}(J_i-J_j) \right].
\end{eqnarray}
  The solution of the system of $n$ equations 
\begin{eqnarray}
\label{partial1}
\frac{\partial F}{\partial J_i}= 2 (1-\lambda) \left[ R_{ii}J_i+  \sum_{j}  R_{ij}(J_i-J_j) \right]+ \nonumber \\
+\lambda \sum_{j} E_{ij}=0,
\end{eqnarray}
 that satisfies the  condition  (\ref{enconv1}) is represented by the point 
in  the  $n$-dimensional liner  space $\{J_i\}$ in which the function  $d_iS/dt$ exhibits  extremum. The value of  
 Lagrange's multiplier $\lambda$   is uniquely determined by the  system of  equations (\ref{partial1}) and   eq. (\ref{enconv1}).

 The standard, rather tedious, procedure of determining  $\lambda$ can be avoided in the following way. 
Let us multiply eq. (\ref{partial1}) with $J_i$ and sum over all mesh currents, i.e.
\begin{eqnarray}
\label{back}
 2 (1-\lambda)\left[ \sum_i R_{ii}J_i^2+  \sum_{ij}  R_{ij}(J_i-J_j)J_i \right]+ \nonumber \\
+\lambda \sum_{ij} E_{ij}J_i=0.
\end{eqnarray}
Due to the symmetry relations (\ref{anti}) and  (\ref{comu}) we have,
\begin{equation}
\label{EE}
\sum_{ij} E_{ij}J_i = - \sum_{ij} E_{ij}J_j,
\end{equation}
and 
\begin{equation}
\label{RR}
\sum_{ij}  R_{ij}J_i^2=\frac{1}{2}\sum_{ij}  R_{ij}(J_i^2+J_j^2).
\end{equation}
 By making use   of eqs. (\ref{EE}) and (\ref{RR}) the  eq.(\ref{back}) becomes,
\begin{eqnarray}
\label{back2}
 2 (1-\lambda)\left[ \sum_i R_{ii}J_i^2+ \frac{1}{2} \sum_{ij}  R_{ij}(J_i-J_j)^2 \right]+ \nonumber \\
+\lambda \left[ \sum_i E_{ii}J_i+\frac{1}{2} \sum_{ij} E_{ij}(J_i-J_j)\right]=0.
\end{eqnarray}
 Substituting the term  linear in mesh currents  in eq. (\ref{back2}) by the term quadratic in mesh currents 
 according to eq. (\ref{enconv1}) we get 
\begin{equation}
\label{labda2}
(2-\lambda)\left[ \sum_i R_{ii}J_i^2+ \frac{1}{2} \sum_{ij}  R_{ij}(J_i-J_j)^2 \right]=0,
\end{equation}
i.e. $\lambda=2$.
 One easily finds that the  system of equations (\ref{partial1}) for $ \lambda =2$ is just the system of 
linear  equations (\ref{lin}), which expresses  the  loop law.

 Let us 
add that $F$ function for $ \lambda =2$ is the $n$-dimensional  paraboloid put upside-down, i.e. its extreme is 
 maximum. In this way the equivalence between the maximum  entropy production principle  and Kirchhoff's loop law 
 is established. In other words we can say that currents in a linear planar  network distribute themselves so as  to achieve the 
 state of    maximum  entropy production.

\section{Discussion }

  It was Ehrenfest (Enzykl. Math. Wissensch, IV, 2(II) fasc.6, p82, note23, 1912.) who first asked the question if there exists some function 
which, like the entropy in the equilibrium state of an isolated system, achieves 
its extreme value in the stationary non-equilibrium state. 

 The minimum  entropy 
production theorem, attributed to Prigogine \cite{prigogine}, identifies the  entropy production 
due to irreversible processes in the system, as  a physical function which becomes 
extremal in a stationary state. We agree with Jaynes´ opinion  \cite{kirchhoff} that Prigogine's 
theorem is in fact the definition of the very special stationary state with 
zero induced flux, known as the static head state in electrical engineering \cite{lePage} and in
bioenergetics \cite{caplan}. It is the  quasi-equilibrium state that can be established close to 
equilibrium in the case of linear relationship between forces and fluxes, 
when induced force is left free to seek its maximal possible value. Energy 
conversion then stops at the level of driving force and corresponding flux, 
because induced flux vanishes and cannot be used to perform any work.
One cannot obtain phenomenological equations, such as those found in references \cite{onsagerI,onsagerII}  
 or eq. (\ref{lin}), from Prigogine's theorem, because that theorem does not have 
 power of a principle.
 On the other hand phenomenological equations, such as eq. (\ref{lin}), can 
  be derived
from  the maximum entropy production principle.

For the one-loop network the maximum entropy production principle holds as well. 
 We leave for a reader to repeat our calculations from equations (\ref{stand}) to (\ref{labda2})
 in the case of the one-loop network. The claim that entropy production is minimal 
 in the stationary state of the one-loop network \cite{kondepudi} is wrong on 
 two accounts. Firstly, authors \cite{kondepudi} have presented potential drops 
 as thermodynamical forces, i.e. they make no distinction between sources and loads 
 and introduce artifical "forces" and "currents" in the entropy production  expression. 
 In the stationary state the charge conservation  law requires  that only  one 
 current flows in the one-loop network, which must be driven by single thermodynamic 
 force (an algebraic sum of all EMF-s in the loop). Secondly, the statement about 
 entropy production minimization in ref. \cite{kondepudi} implies that minimum entropy production 
 theorem \cite{prigogine} holds for a such network. This is not possible, because a minimum 
 of two thermodynamic forces and corresponding currents and the condition of vanishing induced 
 current (the static head non-equilibrium condition) must exist for the minimum entropy 
 production theorem to hold.

  An arbitrary network with no sources of electromotive forces (EMF) was 
first considered with respect to the   entropy production by Kirchhoff  \cite{kirchhoff} , Maxwell \cite{maxwell} ,
and Jeans (see \S 357. in reference \cite{jeans}). Kirchhoff postulated variational theorem  \cite{kirchhoff}, which states that
in an arbitrary volume with fixed surface potential the currents distribute
themselves so as to achieve the state of   minimum  entropy production. Since 
this example contains no sources of EMF, it represents a continuous extension 
of discrete network with no sources which was considered by Maxwell \cite{maxwell} and 
Jeans  (see \S 357. in reference  \cite{jeans}). These authors came to the same conclusion that  minimum  entropy 
production governs current distribution, because they did not consider the 
energy conservation principle.

  When EMF-s and the  energy conservation principle are taken into account then 
the principle of the maximum overall  entropy production in the network leads to 
Kirchhoff´s loop law and accordingly to observed stationary current distribution 
as demonstrated in this paper. Let us note that Jeans considered an  arbitrary,
 not only planar,  electric 
network with EMF-s too. The   maximum  entropy production principle is 
implicitly present in his theorem (see \S 358. in reference \cite{jeans}). 
 This means that the current distribution in non-planar networks is also determined by the 
 maximum entropy production principle.
 Jeans did not define independent currents
in an arbitrary network, as we have done by introducing mesh currents in full
agreement with charge conservation law. Therefore, it is relatively hard to 
follow his derivation.

  Our macroscopic approach is very similar to that of Onsager who considered 
 the related problem of heat conduction in an anisotropic crystal \cite{onsagerI}. We shall
show in a separate paper (in preparation) that an equivalence exists between the 
principle of least dissipation of energy \cite{onsagerI,onsagerII} and the maximum  entropy 
production principle. Onsager was aware of the fact that the stationary state of heat 
conduction is in fact the state of   maximum  entropy production (see the sixth 
section in the reference \cite{onsagerI}). The system of equations that describes stationary 
processes can be inferred from the maximum  entropy production principle both 
in the case of heat conduction and in the case of linear planar electric network.
The advantage of the maximum  entropy production principle is that it offers a  better
physical insight through the explicit use of the energy conservation law in the 
 conditional extremum problem. Rate of work done must be equal to heat produced in a 
stationary state.
 The dissipation function originally introduced by Lord 
Rayleigh \cite{rayleigh} and also  used by Onsager \cite{onsagerI}, obscures the distinction between
 energy sources and passive elements in the system, between the  rate of work done
and heat produced.

  Beside macroscopic formulation of the maximum  entropy production principle there 
exists the microscopic formulation too. It is due to Kohler \cite{kohler} who demonstrated, 
starting from the Boltzmann transport equation, that fluxes in the stationary 
state of gases distribute themselves in order to achieve the state of maximum 
entropy production. Ziman \cite{ziman} extended his work to free electron system in solids 
and demonstrated how this principle can be used in obtaining more accurate solutions
of  the Boltzmann transport equation. Recently, Dewar \cite{dewar} has shown, applying Jaynes´
information theory formalism of statistical mechanics \cite{jaynesI,jaynesII}, 
that stationary states are characterized by  maximum  entropy production.

  To conclude, it is our belief that the principle of   maximum  entropy production 
is valid in linear non-equilibrium thermodynamics, at least for stationary 
processes. In other words we are convinced that it is just the principle Ehrenfest
was looking for. We stressed in this paper that in each situation 
(linear planar electric network in this paper) this principle must be applied as 
the conditional extremum problem  so that energy conservation law holds. Two of 
us have shown recently that biochemical cycle kinetics close to equilibrium state can
be described by an analogue electrical circuit \cite{zupanovic}, and that modeling of photosynthesis \cite{juretic}
can  also be  done by using the maximum  entropy production principle. These and other older 
\cite{old} and more 
recent \cite{dewar,recent} results using this principle 
 show that the present derivation of Kirchhoff's loop law in this paper is
likely to be only the beginning of the widespread use of this principle in many different
scientific disciplines.

\begin{acknowledgments}
The present work is supported by Croatian Ministry of Science and Technology, project no. 0177163 to D.J. 
\end{acknowledgments}

\end{document}